# Exploring User Perceptions of Virtual Reality Scene Design in Metaverse Learning Environments

Rahatara Ferdousi
University of Ottawa
Ottawa, Canada
rferd068@uottawa.ca

Mohd Faisal
University of Ottawa
Ottawa, Canada
mmohd055@uottawa.ca

Fedwa Laamarti
Mohamed bin Zayed University of Artificial Intelligence, UAE
University of Ottawa, Ottawa, Canada
fedwa.laamarti@mbzuai.ac.ae

Chunsheng Yang
Carleton University
Ottawa, Canada
chunsheng.yang@carleton.ca

Abdulmotaleb El Saddik
Mohamed bin Zayed University of Artificial Intelligence, UAE
University of Ottawa Ottawa, Canada
elsaddik@uottawa.ca

*Abstract*—Metaverse learning environments allow for a seamless and intuitive transition between activities compared to Virtual Reality (VR) learning environments, due to their intercon-nected design. The design of VR scenes is important for creating effective learning experiences in the Metaverse. However, there is limited research on the impact of different design elements on user's learning experiences in VR scenes. To address this, a study was conducted with 16 participants who interacted with two VR scenes, each with varying design elements such as style, color, texture, object, and background, while watching a short tutorial. Participant rankings of the scenes for learning were obtained using a seven-point Likert scale, and the Mann-Whitney U test was used to validate differences in preference between the scenes. The results showed a significant difference in preference between the scenes. Further analysis using the NASA TLX questionnaire was conducted to examine the impact of this difference on cognitive load, and participant feedback was also considered. The study emphasizes the importance of careful VR scene design to improve the user's learning experience.

## I. INTRODUCTION

Metaverse has gained attention as a new VR learning environment due its more immersive and interactive design [1]. Similar to VR learning environments, Metaverse learning environments also provide immersive and engaging educational experiences. However, the key difference is that Metaverse learning environments offer a more seamless and intuitive tran-sition between activities due to their interconnected persistent space design [2]. For instance, in a VR learning environment, it is certainly possible to navigate from one lecture space to another, but it may not be as intuitive or seamless as it is in a Metaverse learning environment [3].

Usually, a VR learning environment consists of avatars, multimedia content, multimodal interaction, and VR scenes [4]. Digital twins [5], the virtual representation of avatars [6], multimedia content [7], and immersive interaction [8] for VR learning experience have already been studied in the literature. Designing an effective learning environment requires an un-derstanding of how design elements affect learning outcomes [9]. However, limited research has been conducted on the effect of various design elements, such as style, color, texture, object, and background, on the user's learning experience in VR scenes as illustrated in Fig. 1. Therefore, in this study, we aimed to address the following research questions-RQ1: How do differently designed VR scenes impact the cognitive load in a Metaverse learning environment? RQ2: Do other factors like user's existing vision issues, have an influence on the learning outcome? The contribution of this study is the identification of the significant impact of design elements and existing vision issues of the user's on the learning experience in VR scenes, which can help in the development of future Metaverse learning environments for better user experience.

We conducted a study with 16 participants, who interacted with two VR scenes, each with different design elements (e.g., color, background, texture, etc). while watching a short tutorial. A seven-point Likert scale has been utilized to obtain participant rankings of the scenes for learning, and the Mann-Whitney U test was used to validate differences in preference between the scenes. The results demonstrated a significant difference in preference between the scenes. Further analysis using the NASA TLX questionnaire was conducted to examine the impact of this difference on cognitive load, and participant feedback was also considered for better understanding. Our findings emphasize the importance of careful VR scene design to improve user experience in Metaverse learning environ-ments. It also highlights the need to consider the impact of user's vision issues on the learning experience.

For the immersive experience, we used Meta Quest 2 (formerly known as Oculus Quest) headset with Spatial app [1]. We used the NASA RTLX scores [10] to compare the six dimensions of the cognitive load. We found that different VR scenes and existing vision issues cause various levels of distraction and cognitive demand among the participants.

---
[1] https://www.spatial.io/

The subsequent sections of the paper are structured as follows: Section II presents an overview of VR scenes and existing literature. Section III elaborates on our research procedure, study design, participant details, and other relevant aspects. The results and findings obtained from our investigation are discussed in Section IV. Finally, we conclude this study by providing future directions in Section V.

## II. RELATED WORK

This section presents existing research focusing on the design elements of VR learning environments and the role of those elements in the learning process.

### A. VR Scenes in Metaverse Learning Environment

Metaverse is a platform that enables immersive and interactive learning experiences in a VR environment. VR learning environments leverage the potential of VR technology to create realistic and engaging scenarios for learners [12], [15]. VR scenes have been studied for various applications including virtual tours [17] and simulation-based training [18]. VR scene is a key component of VR learning environments. Virtual spaces populated by 3D models, images, avatars, and humans [4] are some key components of a VR scene. In a Metaverse learning space learners can explore, manipulate, and communicate with various objects and agents in the virtual world [19]. However, the design principles and guidelines for creating effective and engaging VR scenes for learning purposes are still underdeveloped and require further research.

VR educational environments have been categorized into three different categories [1]- i) non-immersive, ii) semi-immersive, and iii) Fully-immersive environments. In this study, we focus on the fully immersive one. The fully immersive environments are usually spaces, where users travel inside the virtual world wearing HMDs and controllers. There are several HMDs like Meta Quest or VIVE PRO which provide full immersion in a learning environment by allowing the users to experience multimedia senses (audio, video, haptics) at the same time. In literature, most of the studies used Oculus Quest due to its portability and low cost.

A common approach in VR research is to create realistic virtual environments that mimic real-world settings. For example, [14] replicated a lecture hall, [16] used paintings to simulate a museum, [15] modeled a shopping scenario, and [12] designed a patient room for training nurses. However, in the context of the Metaverse, more creative and engaging VR designs can be introduced to facilitate learning [1]. Because, the Metaverse spaces may provide an extended reality (XR) experience that goes beyond reality and virtuality [1].

### B. Learning impact in VR environment

In [1], authors found the importance of VR in higher education and the key roles required to design a VR learning platform for multi-stakeholders. This study identified teaching support staff, content creator staff, IT support staff, and digital accessibility staff as important roles to be involved in designing a collaborative VR learning environment.

In [11], authors measured simulator sickness, effectiveness, efficiency, satisfaction, and flow in a VR learning environment. In this study, the participants played a serious game for education [20] which combines gamified knowledge with 3D and VR technology. The authors found immersive VR can enhance learning quality. Similar studies in [6], [15], and [16] measured cognitive task load using the NASA-TLX questionnaire.

In [6], authors applied comment and motion mapping for virtual classrooms and classmates. They found assigning a real learner's time-anchored comments to virtual classmates increases the online learning experience. In [7], authors eval-uated the effectiveness of different learning content represen-tations in virtual environments. They found textual represen-tation in VR is superior to auditory representation in terms of knowledge retention.

Apart from VR classrooms, the cognitive load has also been measured in other contexts like online meetings [13] or shopping [15]. In [8], authors compared six existing Social VR platforms and found inconsistency in communication, spatial navigation, and collaborative editing as the common interaction issues.

### C. Summary

Table I summarizes the application domain, number of participants, and XR tools involved in existing work. Overall, these studies have evaluated the learning experience by mea-suring user preference or cognitive load. However, the impact of variation in the design of VR scenes in a Metaverse learning environment has not been studied. Therefore, this study is conducted to understand it.

## III. METHODS

We designed two different VR Scenes to conduct this study. The design elements of two distinct VR scenes are characterized as follows:

1) Scene-01 is designed as a traditional lecture space (see Figure 1(a)) [2]. To design this scene we utilize a color palette of neutral tones- like beige, brown, and black. The textures in this scene are wooden. This is comple-mented by the presence of traditional objects such as tables, desks, projectors, and chairs, and a campus view background.
2) Scene-02 is designed as a non-traditional lecture space (see Figure 1(b)) [3]. It is characterized by a color palette of bright tones- like pearl white and light blue. The textures here are smooth and clear of any rough or elaborate patterns. Objects in this scene are kept minimal and unique such as shells and pearls, and an aqua reef background.

We selected 1-minute video tutorials on two tech topics including - Introduction to Programming, and Cross-account Protection. The tutorials were shuffled randomly so that the content of tutorials does not affect the learning experience. The study is carried out following the procedure outlined below.

---

[2] https://www.spatial.io/s/Lecture-Room-01-62f9caa877f41c00017d40cf?share=9140823194377408710

[3] https://www.spatial.io/s/Lecture-Room-02-62e0bb1a4a19a000014f390c?share=4699270504055923911

TABLE I: Related work

| Paper | Application Domain | Evaluation Parameter | Participant Number | XR Tools 3D/VR/AR/XR |
|---|---|---|---|---|
| [1] | Workshop in VR and web tools | key roles required for designing a VR learning platform | 18 | Spatial for watching tutorials; Miro and Zoom for brainstorming. Oculus Quest-1 headset |
| [6] | Virtual Classrooms and Classmates | Learning Outcome, Social interactivity and focus attention | 100 | Desktop VR designed in Unity |
| [7] | Educational VR environments | EEG signals Learning outcome, self-reported cognitive load, and Visual attention measures. | 78 | HTC VIVE. |
| [8] | Online Meeting | Comfort and Discomfort during interaction | 17 | Oculus headset with 6 platforms including- |
| [11] | Serious games for affine transformations | Simulator sickness ; Effectiveness; Efficiency; Satisfaction and the flow. | 11 | HTC Vive. |
| [12] | VR learning platform for nurses | Ease of use and the usefulness of VR simulation programs for training nurses | 60 | Oculus Go2 |
| [13] | Social VR | No measurements done | 20-25 | Mozilla hub and HMDs |
| [14] | Immersive classroom | Gaze hit (movement), Spatial presence (action), Temporal demand (time), preferences of representation | 12 | A prototype of Classroom setting with Smart glass. |
| [15] | XR Shopping | Cognitive Workload | 165 | Shop model with Microsoft HoloLens |
| [16] | Reading information in VR | Usability, Cognitive workload, Presence | 18 | Museum gallery with two different representations of texts in 3 different locations. HTC with wifi Adapter. |

1) We welcomed the participants into the experiment room and asked them to complete the registration by signing an informed consent form. We provided each participant with a participant ID and the order of the lecture spaces to be visited. This is because we randomized and shuffled the order of the testing conditions to minimize the order effect.
2) After registration, we requested the participants to in-form us anytime if they feel any type of vision condition - like blurry vision or nausea. We described the flexibil-ity of the experiment that they can take off the headset and end the experiment anytime if they encounter any vision related issues.
3) Then we described the tasks to be performed step-by-step. For this purpose, we demonstrated a pre-recorded one-minute training video. After that, we trained the participants to use Meta Quest 2 and navigation in the spatial app. Furthermore, we showed the participants how to adjust the headset according to their head size. Overall it took around 5 minutes per participant.
4) After introductory training, each participant completed watching a one-minute video tutorial in the first lecture space on their list. After completing one set of tasks for one lecture space, they answered the NASA TLX questionnaire for that particular environment.
5) Finally, the participants finished testing all two test cases. After that, they completed the comparative rank-ing and open feedback form to share their overall experience.

A. Study Design

The study design involved an exploratory investigation of the design element of scenes as the independent variable and with cognitive load and user preferences as the dependent variables. The following section outlines the methodology employed in this study 1.

B. Participants

We recruited 16 participants with diverse VR expertise, gender balance, and an average age of 25.6 years. 62.5% were somewhat familiar with VR, 6.25% were experts, and 32.5% were not at all familiar. 37.5% had myopia, 6.25% had migraines, 6.25% had hypermetropia, and 50% had no vision-related issues.

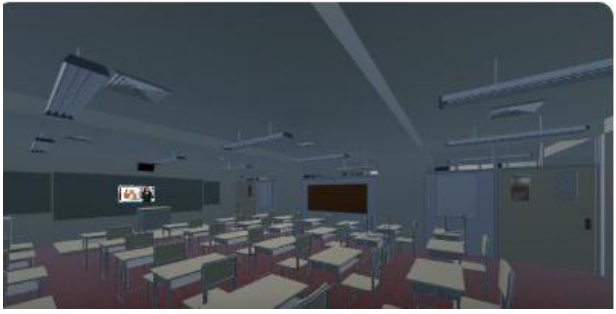

(a) Scene-01: A traditional VR Scene with a campus view in the background. The scene features common objects such as tables, desks, projectors, and chairs.

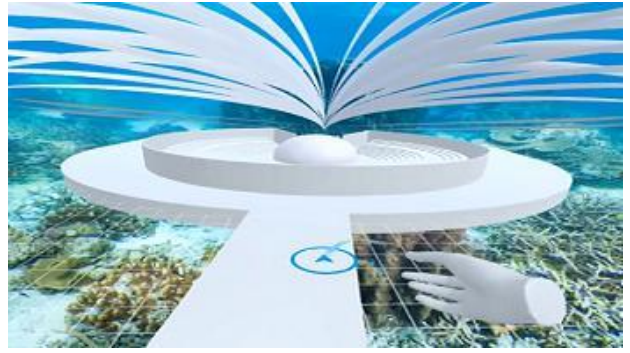

(b) Scene-02: A non-traditional VR Scene, uniquely designed with an aqua reef view in the background. The scene features minimal and common shapes forming a shell and pearls.

Fig. 1: Two different test conditions for this study.

### C. Test Conditions

To investigate the impact of design elements on the learning experience in the Metaverse, we compared two VR scenes, as depicted in Fig. 1. The design elements of each scene are differentiated in detail in Table II.

TABLE II: Design Elements of the VR Scenes

| Design Element | Scene-01 | Scene-02 |
|---|---|---|
| Style | Traditional | Unique and minimalistic |
| Colors | Neutral (beige, brown, black) | Pearl white, light blue |
| Textures | Wooden textures | Smooth surfaces |
| Objects | Traditional (tables, desks, projectors, chairs) | Minimalistic shapes (shell, pearls) |
| Background | Campus view | Aqua reef view |

### D. Materials and Apparatus

We used a Meta Quest 2 headset of 256 GB capacity with controllers and a rubber band for users wearing glasses. The VR lecture spaces were designed with Blender and lecture videos were uploaded to the Spatial. We utilized the original NASA TLX questionnaire to assess the cognitive work load for each scene.

### E. Tasks

For the cognitive load task, participants watched 1-minute video tutorials on two general tech topics in each of the two VR scenes, followed by a series of VR interactions such as navigation, visual and auditory experiences, zooming in and out, pointing, clicking, and moving inside the environment. The tutorials were shuffled randomly to avoid order effects.

### F. Procedure

Participants provided informed consent and were given a participant ID and a randomized order of VR lecture spaces. After a brief training session, participants watched a one-minute video tutorial in each of the two VR lecture spaces as depicted in Fig. 2. They answered the NASA TLX questionnaire for each environment. After completing the assigned task in the two scenes, participants finished the comparative ranking and open feedback form to share their scene preferences and their overall experience. Participants were free to stop at any time if they experienced any vision related issues, and the entire process took approximately 15-20 minutes per participant.

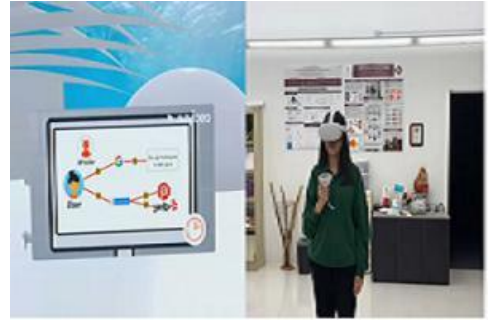

Fig. 2: A participant watching tutorial in Scene-02

## IV. RESULTS AND DISCUSSION

In this section, we present our findings to address RQ1 and RQ2 from our quantitative and qualitative assessments. The scene ranking and Raw NASA TLX score and Man-Whitney answer RQ1 by finding the impact of design elements of VR scenes on the learning experience. The relationship of these scores is analyzed with existing vision issues that helped answer RQ2.

Scene Ranking. The study utilized a seven-point Likert scale to obtain participant rankings of VR scenes for learning. The objective was to validate whether there is a preference among participants for differently designed scenes. The null hypothesis stating that there is no significant difference in preference between Scene-1 and Scene-2 was validated using the Mann-Whitney U test at a significance level of 0.05. The descriptive statistics for the results of the Mann–Whitney U test are presented in TableIII.The obtained p-value of 0.009729 was significantly lower than the significance level. Hence, providing strong evidence to reject the null hypothesis. Therefore, it can be concluded that there is a significant difference in preference between Scene-1 and Scene-2 among participants at the given significance level of 0.05. Further analysis using Raw NASA TLX was conducted to examine the impact of this difference on cognitive load, and participants' open feedback was also considered to understand the reasons behind the varying scene preferences.

TABLE III: Descriptive statistics and results of Mann-Whithney U test for Scene Preferences

|  | Scene-1 | Scene-2 |
|---|---|---|
| Sample Average | 3.93 | 5.62 |
| Sample Size | 16 | 16 |
| Sample Standard deviation | 2.11 | 1.25 |
| Median | 5 | 6 |
| Skewness | -0.19 | -1.24 |
| Rank | 197 | 331 |
| U value | 195 | 61 |
| Mann-Whitney test p-value | p <0.05 | |

Raw NASA TLX Score. The Raw NASA TLX scores presented in Fig. 3 for Scene-1 and Scene-2 suggest that Scene-2 required less effort, frustration, and workload. The Performance demand score was 71.88 for Scene-1 and 76.88 for Scene-2 indicating that individuals found their performance better in Scene-2 than in Scene-1. To enhance the learning ex-perience of Metaverse learning environments, designers should strive to decrease the mental and physical effort needed by learners, reduce frustration levels, and optimize time pressure. Furthermore, it is essential to tailor the learning experience according to individual needs. These results recommend that a minimalist design with plain textures, bright colors, and fewer objects (as Scene-02) could be an effective way to improve the user experience of Metaverse learning environments.

Impact of Vision Related Issues Table IV demonstrates the Raw NASA TLX results for participants both with and without vision issues in Scene-1 and Scene-2. This data helps to comprehend how vision issues influence cognitive load when learning in contrasting virtual reality scenes. The findings suggest vision issues significantly influenced the participants' learning experience in virtual reality. Users with existing vi-sion related issues demonstrated a higher overall cognitive load in both scenes compared to those without any vision prblem. However, the study also revealed that a minimalist VR learning space, like scene-2, helped reduce the overall cognitive load and led to improved user performance, as indicated in Table IV. These results underscore the importance of considering individual vision issues and designing VR environments that promote optimal learning outcomes for all users.

This suggests that the careful design of VR scenes can help to lower the impact of vision issues on the learning experience. Consequently, when creating Metaverse learning environments, it is vital to take vision related issues into account and design scenes that lessen their effect, enabling all learners to gain a favorable learning experience.

Participants Feedback At the end of each experiment, participants gave feedback about their overall learning ex-periences in both Scenes. Several participants found Scene-

TABLE IV: Raw NASA TLX scores for Scene-1 and Scene-2 for both with and without Vision Issues

|  | With Vision Issue | | Without Vision Issue | |
|---|---|---|---|---|
|  | Scene-1 | Scene-2 | Scene-1 | Scene-2 |
| Overall Score | 47.92 | 44.17 | 46.25 | 40.83 |
| Mental | 51.11 | 42.22 | 30.00 | 28.57 |
| Physical | 45.56 | 43.33 | 32.14 | 29.00 |
| Temporal | 55.83 | 47.81 | 48.57 | 47.50 |
| Performance | 63.33 | 85 | 73.57 | 77.14 |
| Effort | 51.11 | 40.56 | 35.83 | 33.57 |
| Frustration | 52.22 | 25 | 19.29 | 20.71 |

02 easier to concentrate on due to its color contrast and plain background. In one of the participant's words- "I found the blue and white color contrast very eye soothing and the underwater environment felt calm and quiet (P1, Scene-02)". Users with vision related issue highlight the importance of considering individual vision related issue while visualizing content in a Metaverse Space. It's evident that for people with myopia, the intense visuals and prolonged use of the headset can lead to discomfort, potential vision issues, and even headaches. One participant with a vision related issue (myopia) stated that "The visuals were quite overwhelming, and after a while with the headset on, I started to have vision changes and a mild headache; it was like everything had been zoomed out (P6, Scene-01)". Therefore, while designing scenes, it is necessary to take the potential impact of vision related issues and minimize it, allowing all learners to have an optimal learning experience.

A. Limitations

This study shows how design elements of a VR scene impact the cognitive load while learning in the Metaverse learning environment. Nonetheless, some limitations exist. These include a small number of participants, leading us to use the Raw NASA TLX score instead of the weighted version. Additionally, the short duration of the video tutorial may impact the understanding of learning impact. However, to understand the learning impact better, lectures of real-life class duration would be better; which we target to achieve in the succeeding part of this study.

V. CONCLUSION AND FUTURE WORK

This study aims to investigate the impact of design elements in VR scenes on the Metaverse learning environment. Our findings indicate that VR design and relevant factors such as color combination, background, texture, and the style of VR scenes significantly impact the user's learning experience. We also discovered that the existing vision problem can cause distractions and affect the learning process. Based on our results, we recommend allowing users to choose VR scenes from multiple options as per their preference, which could lead to improved learning experiences. In addition, other learning elements, such as the mode of content delivery, and the design elements must be considered while designing a Metaverse Learning Space.

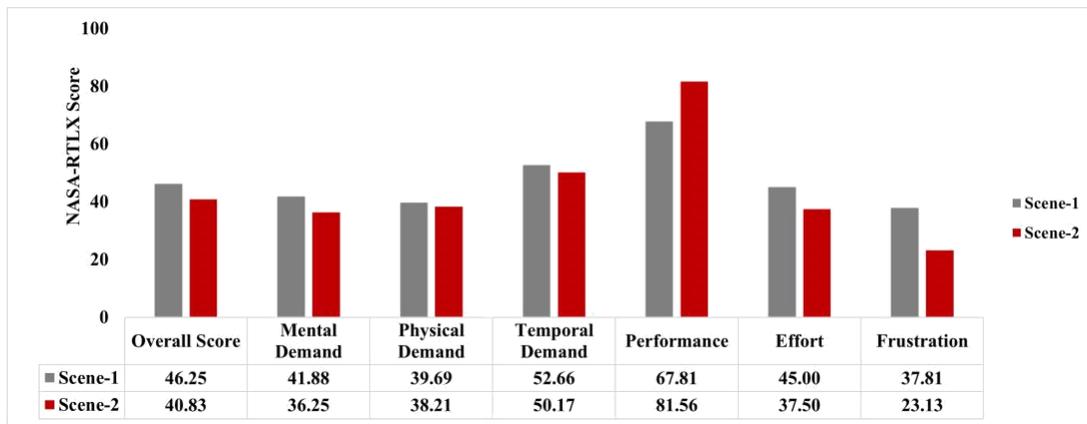

Fig. 3: Comparison of Six Dimensions of Cognitive Workload in Scene-01 and Scene-02